\documentclass{elsart}
\usepackage{natbib}
\usepackage{graphics}
\usepackage{epsfig}
\usepackage{amssymb}

\newcommand{\dvecx}{{\ifmmode d^3 \vec{x} \else $d^3 \vec{x}$\fi}}
\newcommand{\vecx}{{\ifmmode \vec{x} \else $\vec{x}$\fi}}
\newcommand{\tpow}[1]{\ifmmode \times 10^{#1} \else $\times 10^{#1}$ \fi}

\newcommand{\mchi}{{\ifmmode m_{\chi} \else $m_{\chi}$\fi}}
\newcommand{\Msol}{{\ifmmode M_{\odot} \else $M_{\odot}$\fi}}
\newcommand{\Rsol}{{\ifmmode R_{\odot} \else $R_{\odot}$\fi}}
\newcommand{\xsol}{{\ifmmode \vec{x}_{\odot} \else $\vec{x}_{\odot}$\fi}}
\newcommand{\rhosol}{{\ifmmode \rho_{\odot} \else $\rho_{\odot}$\fi}}
\newcommand{\fsol}{{\ifmmode f_{\odot} \else $f_{\odot}$\fi}}

\newcommand{\gtilde}{{\ifmmode \tilde{\cal G} \else $\tilde{{\cal G}}$\fi}}

\newcommand{\pbar}{{\ifmmode \bar{\rm p} \else $\bar{{\rm p}}$\fi}}
\newcommand{\pbars}{{\ifmmode \bar{\rm p}{\rm 's} \else 
    $\bar{{\rm p}}{\rm 's}$\fi}}
\newcommand{\pos}{{\ifmmode {\rm e}^+ \else ${\rm e}^+$\fi}}
\newcommand{\poss}{{\ifmmode {\rm e}^{+} {\rm 's} \else 
    ${\rm e}^{+} {\rm 's}$\fi}}

\begin{document}
\begin{frontmatter}

\title{Anti-proton and positron Cosmic Rays from Dark Matter annihilation 
  around Intermediate Mass Black Holes}

\author[label1,label2]{Julien Lavalle}
\address[label1]{Centre de Physique des Particules, CNRS-IN2P3 / Universit\'e 
de la M\'editerran\'ee, 163 avenue de Luminy, 13288 Marseille Cedex 09 -- 
France}
\ead{lavalle@in2p3.fr OR lavalle@to.infn.it}
\thanks[label2]{Current address: Department of Theoretical Physics, University 
of Turin, Via Giuria 1, 10125 Turin - Italy}

\begin{abstract}
Intermediate Mass Black Holes (IMBHs) are candidates to seed the Supermassive 
Black Holes (SMBHs), and some could still wander in the Galaxy. 
In the context of annihilating dark matter (DM), they are expected to drive 
huge annihilation rates, and could therefore significantly enhance the primary 
cosmic rays (CRs) expected from annihilation of the DM of the 
Galactic halo. In this proceeding (the original paper is 
\citealt{brun_etal_07}), we briefly explain the method to derive 
estimates of such exotic contributions to the \pbar\ and \pos\ CR 
spectra, and the associated statistical uncertainties connected to the 
properties of IMBHs. We find boost factors of order $10^4$ to the exotic 
fluxes, but associated with very large statistical uncertainties.
\end{abstract}


\end{frontmatter}

As in other topics in fundamental physics, annihilating dark matter can be 
motivated by some \emph{coincidence} argument, because exotic microscopic 
physics is expected to surge at the electroweak scale: This energy scale gives 
ideal masses as well as good interaction strengths to naturally get the 
correct abundance of WIMPs today, provided that they are thermally produced 
in the early Universe, without matter-antimatter asymmetry (see the very nice 
lecture by~\citealt{review_dm_murayama_07}).
In the original paper~\citep{brun_etal_07}, we have studied the possibility 
that IMBHs could be the sources of a copious amount of DM annihilation 
products, more precisely \pbars\ and \poss. Such sources have been 
proposed by~\cite{imbh_gamma_bertone_etal_05} in the 
frame of indirect detection of DM with $\gamma$-rays, and further 
extended to neutrinos~\citep{imbh_nu_bertone_06}. Using the 
method developed by~\cite{boost_method_lavalle_etal_07}, 
we have computed how IMBHs may \emph{boost} the exotic 
production of \poss\ and \pbars, compared to considering only 
annihilation of the DM of the Galactic smooth halo (study 
complementary to ~\citealt{bringmann_salati_07}). In this 
proceeding, we shortly present the method to estimate the 
boost factors to the fluxes for both species and the 
corresponding variances associated with the statistical properties of IMBHs. 
The exhaustive study with some predictions for benchmark WIMP models 
(supersymmetry or extra-dimensions) are available in the original paper. We 
concentrate on scenario B 
of~\cite{imbh_gamma_bertone_etal_05} for the formation of IMBHs during the 
matter era. In this scenario, IMBHs form out of cold gas in early-collapsed 
DM halos and are typified by a mass scale of order $10^6\Msol$. 
During their formation, if adiabatic, IMBHs modify the DM density profiles of 
their host halos, so that spikes appear at the centres. Starting with an 
$r^{-\gamma}$ spherical density profile, the final logarithmic slope due to 
the adiabatic compression of dark matter is $\gamma_{\rm fin} = 
(9-2\gamma)/(4-\gamma)$. For instance, taking a profile initially with 
$\gamma = 1$ leads to $\gamma_{\rm fin} = 7/3$, which stands inside the spike 
radius $r_{\rm sp}$. Whatever the DM profile surrounding an IMBH, it is useful 
to define a quantity proportional to the corresponding WIMP annihilation rate, 
the \emph{effective annihilation volume}, as follows:
$\xi \equiv \int_{\rm BH} d^3\vec{x} \left( \frac{\rho_{\rm BH}}{\rho_\odot}
\right)^2\;$,
where $\rho_{\rm BH}$ is the DM density profile around the black 
hole, and, $\rho_\odot$, the local dark matter density, allows for a 
normalisation to the local annihilation rate. Such an effective volume 
is the one the DM stuck to the IMBH would have if it were diluted 
down to the local density $\rho_\odot$.
Note that $\xi$ is sensitive to the inner cut-off radius $r_{\rm c}$ below 
which the spike density saturates, which is usually set by equating the 
gravitational infall rate with the WIMP annihilation rate. This cut-off 
is associated with the maximum value of the accreted DM density 
$\rho_{\rm max}=\rho(r_{\rm c})\simeq m_{\chi}/(\langle 
\sigma_{\rm ann}v\rangle \tau)$, where $\tau \simeq 10$ Gyr is the typical 
IMBH formation timescale. If computed inside $r_{\rm sp}$, 
$\xi \propto (m_{\chi}/\langle \sigma_{\rm ann}v \rangle)^{5/7}$, and only 
slightly depends on the WIMP model.
Considering IMBHs forming in early-collapsing halos ($r^{-1}$ profiles), 
\cite{imbh_gamma_bertone_etal_05} performed simulations to derive the 
evolution of their DM spikes in the Galaxy. From their 
results, they extracted the statistical properties of the 
surviving spikes, i.e. their spatial distribution in the Galaxy, and the 
surrounding DM profile properties. The total number of IMBHs is 
small, $\sim 100$ within a galactocentric radius of 500 kpc. Nonetheless, the 
probability to find an IMBH increases when going toward the Galactic 
Centre, and the Earth is located at an interesting place from an 
observational point of view. The whole phase space of IMBH 
properties is thus portrayed by the spatial distribution 
$d{\cal P}_V/dV$ of objects in the Galaxy and the distribution of their 
effective annihilation volumes $\xi$, $d{\cal P}_\xi / d\xi$, both extracted 
from these simulations, that we used. For generic WIMPs, the typical mean 
value $\langle \xi \rangle \sim 10^6$ kpc$^3$, which is huge.
%
The CR propagation modelling is a key ingredient for this kind of 
studies. Here, we adopt a slab diffusion zone, featured by its radial 
extension that we fix to $R_{\rm slab}=30$ kpc, and by its half-thickness 
$L$, $4$ kpc here. CRs are confined within the slab, which translates 
to Dirichlet boundary conditions for the diffusion equation (the CR 
number density vanishes on the borders). For the transport processes, we 
take a spatial independent diffusion coefficient 
$K(E)=\beta K_0{\cal R}^{\delta}$ (where ${\cal R}=pc/Ze$ is the rigidity) and 
a constant wind $V_{\rm conv}$ directed outwards along the vertical axis. Such 
a configuration is reminiscent from the \emph{medium} set of parameters 
provided by~\citealt{2001ApJ...555..585M}, that we choose 
here: $K_0 = 0.0112$ kpc$^2$/Myr, $\delta = 0.7$ and $V_{\rm conv} = 12$ km/s. 
One can easily write and solve the diffusion equations for both \poss\ and 
\pbars\ for this kind of geometry.
For \poss, the main processes that come into play are the energy losses 
(mainly inverse Compton diffusion off CMB or IR photons, and synchrotron 
radiation), and the diffusion on the magnetic turbulences. Disregarding 
the convection process, which is much less efficient than energy losses, 
one can express the typical propagation length for \poss\ as:
$\lambda_{\rm d} \equiv \left(2 K_0 \tau_E 
\left( \frac{\epsilon^{\delta - 1} -
  \epsilon_{S}^{\delta-1}}{1-\delta}\right)\right)^{1/2}$ 
where $\epsilon\equiv E/\{E_0 = 1\;{\rm GeV}\}$. Assuming an infinite 
3D spherical diffusion zone (correct while $\lambda_{\rm d}\lesssim L$), the 
\pos\ propagator is proportional to a Gaussian function of the source 
distance with $\sigma = \lambda_{\rm d}$ ($\lesssim$ few kpc): Sources located 
farther than $\lambda_{\rm d}$ will almost not contribute to the flux at the 
Earth. $\lambda_d$ being a decreasing function of the detected 
energy, the effective volume in which \poss\ propagate increases as they 
loose energy.
The diffusion equation for \pbars\ must include spallation processes and 
wind convection that occur in the thin Galactic disc, so we can not 
use a simple spherical symmetry to derive a global expression. Moreover, 
the energy losses are negligible for this species, which modifies 
significantly the picture that we had for \poss. Nevertheless, it is 
useful to write, as for \poss, the typical propagation length:
$\Lambda_{\rm d} \equiv \frac{K(E)}{V_{\rm conv}}$,
where convection is assumed to dominate over spallation in average, which is 
correct unless at sub-GeV energies. This is quite different from \poss\ 
because this length is an increasing function of energy. 
The picture is therefore reversed, and the propagation volume is much larger 
at higher energy for \pbars\ ($\Lambda_{\rm d}$ reaches 
the size of the diffusion slab at energies of order 10-100 GeV).
We now define a convenient Green function for any CR species, that 
encodes the injected spectrum induced by DM annihilation:
$\gtilde (E,\xsol \leftarrow \vecx) \equiv \int_{E}^{E_{\rm max}} 
dE_S \; {\cal G} (E,\xsol \leftarrow E_S,\vecx)\times \frac{dN_{\rm CR}(E_S)}
{dE_S}$,
where $dN_{\rm CR}/dE_S$ is the injected spectrum at source ($E = E_S$ for 
\pbars).
The Galactic host halo is described by a smooth DM distribution 
$\rho_{\rm sm}$, so that the corresponding primary CR flux reads:
$\phi_{\rm sm}(E) = \frac{v}{4\pi}{\cal S}\int_{\rm halo} \dvecx \; 
\gtilde (E,\xsol \leftarrow \vecx) \left(\frac{\rho_{\rm sm}}{\rhosol}
\right)^2$
where $v$ is the CR velocity and ${\cal S}\equiv \delta \langle 
\sigma_{\rm ann} v\rangle \rhosol^2/(2\mchi^2)$ encodes the main WIMP 
properties\footnote{$\delta = 1/2$ for Majorana particles, 1 otherwise.}.
%
IMBHs can be considered as point-like sources, and the flux due to 
the $i^{\rm th}$ object is merely:
$\phi_{{\rm bh},i} (E) = \frac{v}{4\pi}\times {\cal S} \times \xi_i \times 
\gtilde (E,\xsol \leftarrow \vecx_i)$.
Then, we have to sum over the whole population. We can derive a 
statistical prediction by using the IMBH phase space information, either 
with MC simulations or with a semi-analytic method. The overall IMBH 
contribution is:
$\phi_{\rm bh,tot} (E) = \frac{1}{\cal N}\sum_{i=1}^{\cal N} 
\sum_{j=1}^{N_i} \phi_{i,j}(E)
\rightarrow  \frac{v}{4\pi}\int dN N \frac{d{\cal P}_N(N)}{dN} \times 
\int d\xi \xi \frac{d{\cal P}_\xi(\xi)}{d\xi} \times \int\dvecx 
\gtilde(E,\xsol \leftarrow \vecx) \frac{d{\cal P}_V(\vecx)}{\dvecx}
 =  \langle N \rangle \langle \phi_{\rm bh} \rangle = 
\frac{v}{4\pi}{\cal S} \langle N \rangle \langle \xi \rangle 
\langle \gtilde \rangle$
where the 1$^{\rm st}$ equality is a mere counting (performed e.g. with a 
Monte Carlo | MC), and the 2$^{\rm nd}$ one takes directly the | normalised | 
pdfs into account (this limit is reached for an infinite number ${\cal N}$ 
of MC realisations). Those pdfs characterise the total number $N$ of 
IMBHs wandering in the Galactic halo, $\xi$ and \gtilde, the latter being 
spatially weighted with $d{\cal P}_V/dV$. The last equality gives the same 
quantities in terms of statistical mean values. This assumes no 
correlations between the considered variables, which has been carefully 
checked from the simulation results.
The calculation of the variance of CR fluxes originating from the whole IMBH 
sample $\sigma_{\rm bh,tot}$ is straightforward:
$\frac{\sigma_{\rm bh,tot}^2}{\phi_{\rm bh,tot}^2} = 
\frac{1}{\langle N \rangle} \left(\frac{\sigma_\xi^2}{\langle\xi\rangle^2}+ 
\frac{\sigma_\gtilde^2}{\langle\gtilde\rangle^2} +
\frac{\sigma_\xi^2\sigma_\gtilde^2}{\langle\xi\rangle^2\langle\gtilde\rangle^2}
\right) + \frac{\sigma_N^2}{\langle N\rangle^2}\;$,
where $\sigma_x$ is the variance corresponding to any variable $x$.
The \emph{boost factors} for \poss\ and \pbars\ are the ratios 
of the fluxes originating from a halo populated with IMBHs to those 
calculated for the host smooth halo alone. As the Galaxy mass fraction carried 
by IMBHs is negligible, the effective boost is $B(E) \simeq 1 + 
\phi_{\rm bh,tot}/\phi_{\rm sm}$, which depends on energy.
Estimates of the mean fluxes for \poss\ and \pbars\ and associated 
statistical variances have been performed by using (i) a MC method 
(ii) a semi-analytic calculation based on the method proposed by 
\cite{boost_method_lavalle_etal_07}. Both methods 
agree, and the average boost factors are found to be $\sim 10^4$ for both 
species, but with a very large relative variance, going up to $\sim$ 100\% 
as the CR propagation scale decreases ; those results are very different 
from what is found in the context of standard DM clumpiness, for 
which the enhancement is negligible~\citep{boost_clumps_lavalle_etal_07}. The 
large statistical error found is connected with the small number 
of IMBHs predicted inside the diffusion volume set by the CR propagation 
scale, and more precisely with the distance of the closest object. Should it 
be close, which is statistically unlikely, it would strongly dominate the 
exotic contribution of antimatter, otherwise it would remain almost 
unobservable. The existing measurements of the \pos\ and 
\pbar\ fluxes already provide strong constraints that we will 
translate in limits on the gamma-ray and neutrino production from IMBHs in a 
forthcoming paper.\\
\underline{Acknowledgements}: The author is grateful to his collaborators, 
G.~Bertone, P.~Brun, P.~Salati and R.~Taillet for enlightening 
discussions during this work.

\end{document}